\pdfoutput=1

\documentclass[12pt,a4paper]{article}

\usepackage{ifthen} % for conditional statements
\newboolean{pdflatex}
\setboolean{pdflatex}{true} % False for eps figures 

\newboolean{articletitles}
\setboolean{articletitles}{true} % False removes titles in references

\newboolean{uprightparticles}
\setboolean{uprightparticles}{false} %True for upright particle symbols

\def\paperauthors{LHCb collaboration} 

\def\paperasciititle{Search for the lepton-flavour violating decays B0 to Kstar0 tau mu } 
\def\papertitle{Search for the lepton-flavour violating decays $B^0 \to K^{*0} \tau^\pm \mu^\mp$} % Latex formatted title
\def\paperkeywords{{High Energy Physics}, {LHCb}} % Comma separated list
\def\papercopyright{\the\year\ CERN for the benefit of the LHCb collaboration} 
\def\paperlicence{CC BY 4.0 licence}
\def\paperlicenceurl{https://creativecommons.org/licenses/by/4.0/}

\usepackage[top=1in, bottom=1.25in, left=1in, right=1in]{geometry}

\usepackage{microtype}
\usepackage{lineno}  % for line numbering during review
\usepackage{xspace} % To avoid problems with missing or double spaces after
\usepackage{caption} %these three command get the figure and table captions automatically small

\usepackage{graphicx}  % to include figures (can also use other packages)
\usepackage{color}
\usepackage{colortbl}
\graphicspath{{./figs/}} % Make Latex search fig subdir for figures
\usepackage{amsmath} % Adds a large collection of math symbols
\usepackage{amssymb}
\usepackage{amsfonts}
\usepackage{upgreek} % Adds in support for greek letters in roman typeset

\newcommand*\patchAmsMathEnvironmentForLineno[1]{%
\expandafter\let\csname old#1\expandafter\endcsname\csname #1\endcsname
\expandafter\let\csname oldend#1\expandafter\endcsname\csname
end#1\endcsname
 \renewenvironment{#1}%
   {\linenomath\csname old#1\endcsname}%
   {\csname oldend#1\endcsname\endlinenomath}%
}
\newcommand*\patchBothAmsMathEnvironmentsForLineno[1]{%
  \patchAmsMathEnvironmentForLineno{#1}%
  \patchAmsMathEnvironmentForLineno{#1*}%
}
\AtBeginDocument{%
\patchBothAmsMathEnvironmentsForLineno{equation}%
\patchBothAmsMathEnvironmentsForLineno{align}%
\patchBothAmsMathEnvironmentsForLineno{flalign}%
\patchBothAmsMathEnvironmentsForLineno{alignat}%
\patchBothAmsMathEnvironmentsForLineno{gather}%
\patchBothAmsMathEnvironmentsForLineno{multline}%
\patchBothAmsMathEnvironmentsForLineno{eqnarray}%
}

\usepackage{hyperxmp}

\usepackage[pdftex,
            pdfauthor={\paperauthors},
            pdftitle={\paperasciititle},
            pdfkeywords={\paperkeywords},
            pdfcopyright={Copyright (C) \papercopyright},
            pdflicenseurl={\paperlicenceurl}]{hyperref}
\usepackage[colorinlistoftodos,textsize=scriptsize]{todonotes}

\usepackage[bottom,flushmargin,hang,multiple]{footmisc}

\usepackage[all]{hypcap} % Internal hyperlinks to floats.

\usepackage{xspace} 
\usepackage{upgreek}

\def\lhcb   {\mbox{LHCb}\xspace}

\def\MagUp {\mbox{\em Mag\kern -0.05em Up}\xspace}

\ifthenelse{\boolean{uprightparticles}}%
{

 \def\Pmu         {\ensuremath{\upmu}\xspace}                 
 \def\Pnu         {\ensuremath{\upnu}\xspace}                 
                  
 \def\Ppi         {\ensuremath{\uppi}\xspace}

 \def\Ptau        {\ensuremath{\uptau}\xspace}

 \def\Ppsi        {\ensuremath{\uppsi}\xspace}

 \def\PDelta      {\ensuremath{\Delta}\xspace}                 
 \def\PXi         {\ensuremath{\Xi}\xspace}                 
 \def\PLambda     {\ensuremath{\Lambda}\xspace}                 
 \def\PSigma      {\ensuremath{\Sigma}\xspace}                 
 \def\POmega      {\ensuremath{\Omega}\xspace}                 
 \def\PUpsilon    {\ensuremath{\Upsilon}\xspace}
 \let\oldPi\Pi
 \def\PPi         {\ensuremath{\oldPi}\xspace}

 \def\PB      {\ensuremath{\mathrm{B}}\xspace}                 
                  
 \def\PD      {\ensuremath{\mathrm{D}}\xspace}

 \def\PJ      {\ensuremath{\mathrm{J}}\xspace}                 
 \def\PK      {\ensuremath{\mathrm{K}}\xspace}

 \def\Pb      {\ensuremath{\mathrm{b}}\xspace}                 
 \def\Pc      {\ensuremath{\mathrm{c}}\xspace}

 \def\Pi      {\ensuremath{\mathrm{i}}\xspace}

 \def\Ps      {\ensuremath{\mathrm{s}}\xspace}

 \def\thebaroffset{0.0em}
}
{

 \def\Pmu         {\ensuremath{\mu}\xspace}                 
 \def\Pnu         {\ensuremath{\nu}\xspace}                 
                  
 \def\Ppi         {\ensuremath{\pi}\xspace}

 \def\Ptau        {\ensuremath{\tau}\xspace}

 \def\Ppsi        {\ensuremath{\psi}\xspace}                 
                  
 \mathchardef\PDelta="7101
 \mathchardef\PXi="7104
 \mathchardef\PLambda="7103
 \mathchardef\PSigma="7106
 \mathchardef\POmega="710A
 \mathchardef\PUpsilon="7107
 \mathchardef\PPi="7105
                  
 \def\PB      {\ensuremath{B}\xspace}                 
                  
 \def\PD      {\ensuremath{D}\xspace}

 \def\PJ      {\ensuremath{J}\xspace}                 
 \def\PK      {\ensuremath{K}\xspace}

 \def\Pb      {\ensuremath{b}\xspace}                 
 \def\Pc      {\ensuremath{c}\xspace}

 \def\Pi      {\ensuremath{i}\xspace}

 \def\Ps      {\ensuremath{s}\xspace}

 \def\thebaroffset{0.18em}
}
\newcommand{\offsetoverline}[2][\thebaroffset]{\kern #1\overline{\kern -#1 #2}}%

\makeatletter
\ifcase \@ptsize \relax% 10pt
  \newcommand{\miniscule}{\@setfontsize\miniscule{4}{5}}% \tiny: 5/6
\or% 11pt
  \newcommand{\miniscule}{\@setfontsize\miniscule{5}{6}}% \tiny: 6/7
\or% 12pt
  \newcommand{\miniscule}{\@setfontsize\miniscule{5}{6}}% \tiny: 6/7
\fi
\makeatother

\DeclareRobustCommand{\optbar}[1]{\shortstack{{\miniscule (\rule[.5ex]{1.25em}{.18mm})}
  \\ [-.7ex] $#1$}}

\def\mumu       {{\ensuremath{\Pmu^+\Pmu^-}}\xspace}

\def\taum       {{\ensuremath{\Ptau^-}}\xspace}

\def\neu        {{\ensuremath{\Pnu}}\xspace}

\def\neut       {{\ensuremath{\neu_\tau}}\xspace}

\def\squark    {{\ensuremath{\Ps}}\xspace}

\def\cquark    {{\ensuremath{\Pc}}\xspace}

\def\bquark    {{\ensuremath{\Pb}}\xspace}
\def\bquarkbar {{\ensuremath{\overline \bquark}}\xspace}
\def\bbbar     {{\ensuremath{\bquark\bquarkbar}}\xspace}

\def\pion   {{\ensuremath{\Ppi}}\xspace}
\def\piz    {{\ensuremath{\pion^0}}\xspace}
\def\pip    {{\ensuremath{\pion^+}}\xspace}
\def\pim    {{\ensuremath{\pion^-}}\xspace}

\def\kaon    {{\ensuremath{\PK}}\xspace}
\def\KorKbar {\kern \thebaroffset\optbar{\kern -\thebaroffset \PK}{}\xspace}

\def\Kp      {{\ensuremath{\kaon^+}}\xspace}
\def\Km      {{\ensuremath{\kaon^-}}\xspace}

\def\Kstarz  {{\ensuremath{\kaon^{*0}}}\xspace}

\def\Dbar    {{\ensuremath{\offsetoverline{\PD}}}\xspace}
\def\D       {{\ensuremath{\PD}}\xspace}

\def\DorDbar {\kern \thebaroffset\optbar{\kern -\thebaroffset \PD}\xspace}
\def\Dz      {{\ensuremath{\D^0}}\xspace}
\def\Dzb     {{\ensuremath{\Dbar{}^0}}\xspace}
\def\Dp      {{\ensuremath{\D^+}}\xspace}
\def\Dm      {{\ensuremath{\D^-}}\xspace}

\def\DpDm    {\ensuremath{\Dp {\kern -0.16em \Dm}}\xspace}
\def\Dstar   {{\ensuremath{\D^*}}\xspace}

\def\Dstarz  {{\ensuremath{\D^{*0}}}\xspace}

\def\Ds      {{\ensuremath{\D^+_\squark}}\xspace}

\def\B       {{\ensuremath{\PB}}\xspace}

\def\BorBbar {\kern \thebaroffset\optbar{\kern -\thebaroffset \PB}\xspace}

\def\Bd      {{\ensuremath{\B^0}}\xspace}

\def\BdorBdbar {\kern \thebaroffset\optbar{\kern -\thebaroffset \Bd}\xspace}

\def\Bs      {{\ensuremath{\B^0_\squark}}\xspace}

\def\BsorBsbar {\kern \thebaroffset\optbar{\kern -\thebaroffset \Bs}\xspace}

\def\jpsi     {{\ensuremath{{\PJ\mskip -3mu/\mskip -2mu\Ppsi}}}\xspace}

\def\Y#1S{\ensuremath{\PUpsilon{(#1S)}}\xspace}

\def\LorLbar     {\kern \thebaroffset\optbar{\kern -\thebaroffset \PLambda}\xspace}

\def\BF         {{\ensuremath{\mathcal{B}}}\xspace}
\def\BR         {\BF}

\newcommand{\decay}[2]{\ensuremath{#1\!\to #2}\xspace} 

\def\to                 {\ensuremath{\rightarrow}\xspace}

\def\AT#1     {\ensuremath{A_{\mathrm{T}}^{#1}}\xspace}           % 2

\def\C#1      {\ensuremath{\mathcal{C}_{#1}}\xspace}                       % 9
\def\Cp#1     {\ensuremath{\mathcal{C}_{#1}^{'}}\xspace}                    % 7
\def\Ceff#1   {\ensuremath{\mathcal{C}_{#1}^{\mathrm{(eff)}}}\xspace}        % 9  
\def\Cpeff#1  {\ensuremath{\mathcal{C}_{#1}^{'\mathrm{(eff)}}}\xspace}       % 7
\def\Ope#1    {\ensuremath{\mathcal{O}_{#1}}\xspace}                       % 2
\def\Opep#1   {\ensuremath{\mathcal{O}_{#1}^{'}}\xspace}                    % 7

\newcommand{\nospaceunit}[1]{\ensuremath{\text{#1}}}       
\newcommand{\aunit}[1]{\ensuremath{\text{\,#1}}}       
\newcommand{\tev}{\aunit{Te\kern -0.1em V}\xspace}
\newcommand{\gev}{\aunit{Ge\kern -0.1em V}\xspace}
\newcommand{\mev}{\aunit{Me\kern -0.1em V}\xspace}
\newcommand{\kev}{\aunit{ke\kern -0.1em V}\xspace}
\newcommand{\ev}{\aunit{e\kern -0.1em V}\xspace}
 
\newcommand{\mevc}{\ensuremath{\aunit{Me\kern -0.1em V\!/}c}\xspace}
\newcommand{\gevc}{\ensuremath{\aunit{Ge\kern -0.1em V\!/}c}\xspace}
\newcommand{\mevcc}{\ensuremath{\aunit{Me\kern -0.1em V\!/}c^2}\xspace}
\newcommand{\gevcc}{\ensuremath{\aunit{Ge\kern -0.1em V\!/}c^2}\xspace}
\newcommand{\gevgevcccc}{\ensuremath{\gev^2\!/c^4}\xspace} % for q^2

\def\mm   {\aunit{mm}\xspace}

\def\mum  {\ensuremath{\,\upmu\nospaceunit{m}}\xspace}

\def\fb   {\ensuremath{\aunit{fb}}\xspace}
\def\invfb   {\ensuremath{\fb^{-1}}\xspace}

\def\gsim{{~\raise.15em\hbox{$>$}\kern-.85em
          \lower.35em\hbox{$\sim$}~}\xspace}
\def\lsim{{~\raise.15em\hbox{$<$}\kern-.85em
          \lower.35em\hbox{$\sim$}~}\xspace}

\def\pt         {\ensuremath{p_{\mathrm{T}}}\xspace}

\def\ptot       {\ensuremath{p}\xspace}

\def\evtgen     {\mbox{\textsc{EvtGen}}\xspace}

\def\geant      {\mbox{\textsc{Geant4}}\xspace}

\def\photos     {\mbox{\textsc{Photos}}\xspace}

\def\pythia     {\mbox{\textsc{Pythia}}\xspace}

\def\tell1  {TELL1\xspace}
\def\ukl1   {UKL1\xspace}

\newcommand{\ie}{\mbox{\itshape i.e.}\xspace}

\usepackage{cite} % Allows for ranges in citations
\usepackage{mciteplus}
\def\tauola     {\mbox{\textsc{Tauola}}\xspace}

\mathchardef\mhyphen="2D
\newcommand{\kfolding}{\ensuremath{k\mhyphen\mathrm{folding}\xspace}}
\newcommand{\OC}{\mymath{\decay{\Bd}{\Kstarz\tau^-\mu^+}}}
\newcommand{\SC}{\mymath{\decay{\Bd}{\Kstarz\tau^+\mu^-}}}
\renewcommand{\epsilon}{\varepsilon}
\renewcommand{\theta}{\vartheta}

\newcommand{\mymath}[1]{\ensuremath{#1}\xspace}
\newcommand{\DToKPiPi}{\mymath{\decay{\Dm}{\Kp\pim\pim}}}

\newcommand{\DspToKKPi}{\mymath{\decay{\Ds}{\Kp\Km\pip}}}

\newcommand{\TauMu}{\mymath{\tau\mu}}
\newcommand{\osTauMu}{\mymath{\tau^\pm\mu^\mp}}

\newcommand{\BToKstTauMu}{\mymath{\decay{\Bd}{\Kstarz\TauMu}}}

\newcommand{\KstToKPi}{\mymath{\decay{\Kstarz}{\Kp\pim}}}

\newcommand{\osBToKstTauMu}{\mymath{\decay{\Bd}{\Kstarz\osTauMu}}}

\newcommand{\tauToPiPiPi}{\mymath{\decay{\taum}{\pim\pip\pim\neut}}}
\newcommand{\tauToPiPiPiPiz}{\mymath{\decay{\taum}{\pim\pip\pim\piz\neut}}}

\newcommand{\BdToDDs}{\mymath{\decay{\Bd}{\Dm\Ds}}}

\usepackage{enumitem}
\usepackage{cleveref}
\usepackage{multirow}
\usepackage{placeins}
\usepackage{tikz}
\usetikzlibrary{trees,arrows,automata,shapes}
\usetikzlibrary{decorations.pathmorphing}
\usetikzlibrary{decorations.markings}
\usetikzlibrary{shapes.symbols}
\usetikzlibrary{shapes.geometric}
\usetikzlibrary{shapes.arrows}
\usetikzlibrary{positioning}
\tikzset
{
   vector/.style = {decorate, decoration={snake,amplitude=2pt, segment length=5pt}},
   fermion/.style = {postaction={decorate}, decoration={markings,mark=at position .55 with {\arrow{>}}}},
   fermionbar/.style = {postaction={decorate}, decoration={markings,mark=at position .55 with {\arrow{<}}}},
   gluon/.style = {decorate, decoration={coil,amplitude=3pt, segment length=5pt}},
   scalar/.style = {dashed, postaction={decorate}, decoration={markings,mark=at position .55 with {\arrow{>}}}},
   scalarbar/.style = {dashed, postaction={decorate}, decoration={markings,mark=at position .55 with {\arrow{<}}}}
}
\tikzstyle{information text}=[draw,rounded corners,inner sep=1ex]
\usepackage{longtable} 

\begin{document}

%%%%%%%%%%%%%%%%%%%%%%%%%
%%%%% Title     %%%%%%%%%
%%%%%%%%%%%%%%%%%%%%%%%%%
\renewcommand{\thefootnote}{\fnsymbol{footnote}}
\setcounter{footnote}{1}

%%%%%%%%%%%%%%%%%%%%%%%%%
%%%%%  TITLE PAGE  %%%%%%
%%%%%%%%%%%%%%%%%%%%%%%%%
\begin{titlepage}
\pagenumbering{roman}

% Header ---------------------------------------------------
\vspace*{-1.5cm}
\centerline{\large EUROPEAN ORGANIZATION FOR NUCLEAR RESEARCH (CERN)}
\vspace*{1.5cm}
\noindent
\begin{tabular*}{\linewidth}{lc@{\extracolsep{\fill}}r@{\extracolsep{0pt}}}
\ifthenelse{\boolean{pdflatex}}% Logo format choice
{\vspace*{-1.5cm}\mbox{\!\!\!\includegraphics[width=.14\textwidth]{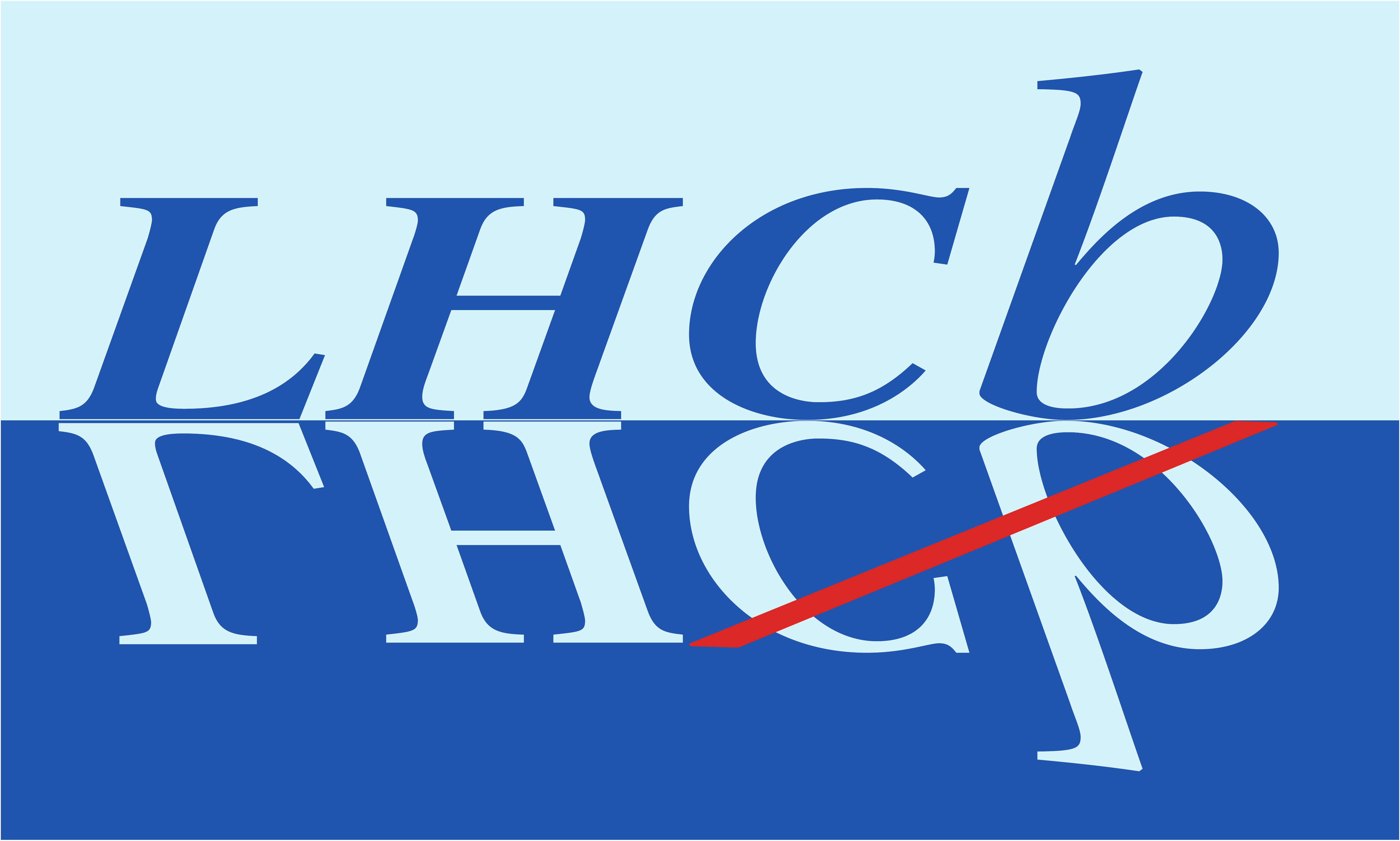}} & &}%
{\vspace*{-1.2cm}\mbox{\!\!\!\includegraphics[width=.12\textwidth]{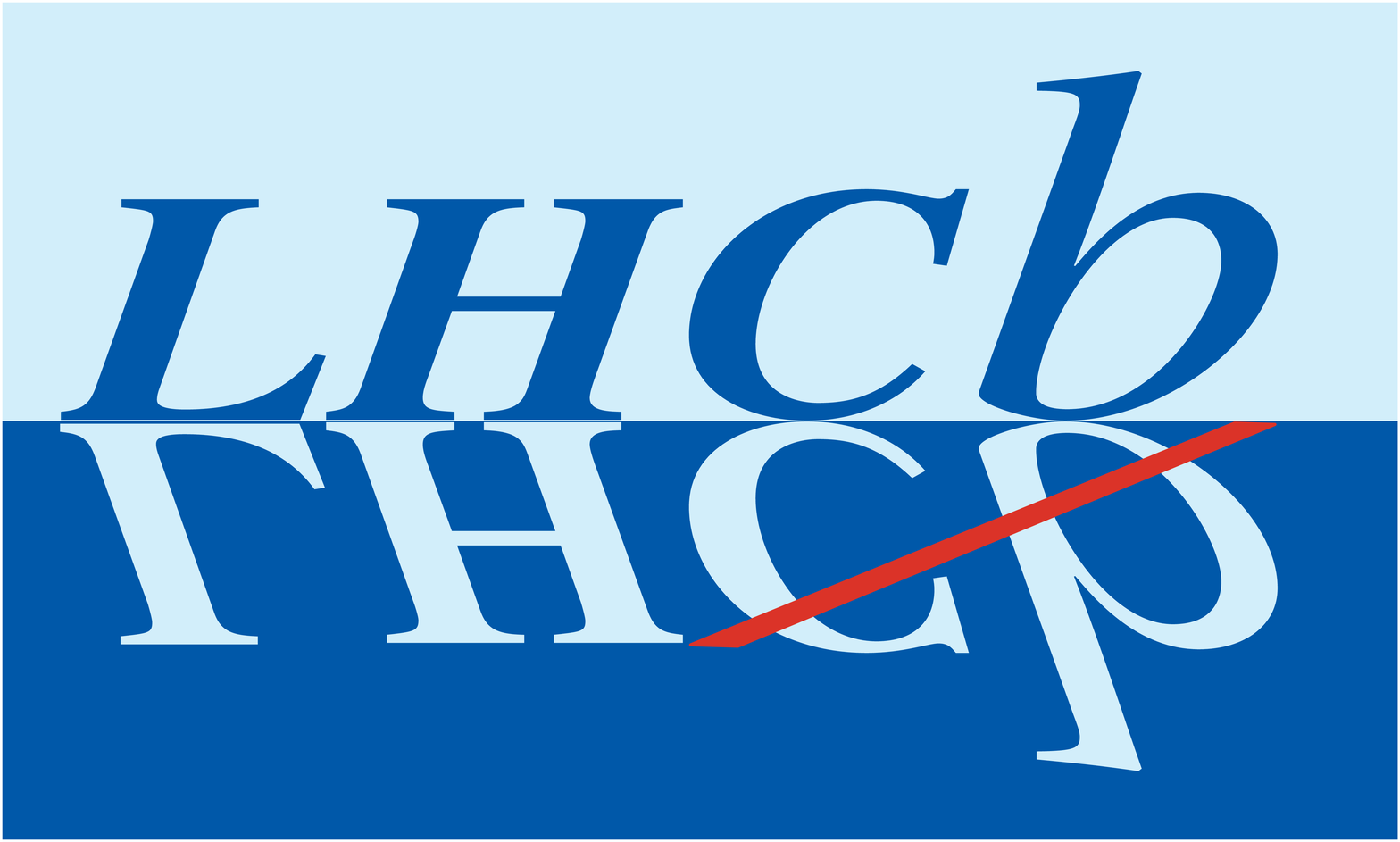}} & &}%
\\
 & & CERN-EP-2022-154 \\  % ID 
 & & LHCb-PAPER-2022-021 \\  % ID 
 %& & \today \\ % Date - Can also hardwire e.g.: 23 March 2010
 & & September 20, 2022 \\ % 
 & & \\
% not in paper \hline
\end{tabular*}

\vspace*{4.0cm}

% Title --------------------------------------------------
{\normalfont\bfseries\boldmath\huge
\begin{center}
% DO NOT EDIT HERE. Instead edit macro in main.tex to keep metadata correct
  \papertitle 
\end{center}
}

\vspace*{2.0cm}

% Authors -------------------------------------------------
\begin{center}
%In the footnote, replace 'paper' by 'Letter' in case of submission to PRL or PLB 
% Edit macro in main.tex to keep metadata correct
\paperauthors
%\footnote{Authors are listed at the end of this paper.}
\end{center}

\vspace{\fill}

% Abstract -----------------------------------------------
\begin{abstract}
  \noindent
  A first search for the lepton-flavour violating decays  $B^0 \to K^{*0} \tau^\pm \mu^\mp$  
is presented.  The analysis is performed using a sample of proton-proton collision data, collected with the LHCb detector at  centre-of-mass energies of 7, 8 and 13 TeV between 2011 and 2018,  corresponding to an integrated luminosity of 9 fb$^{-1}$. No significant signal is observed, and upper limits on the branching fractions are determined to be  ${\cal{B}}(B^0 \to K^{*0} \tau^+ \mu^-)< 1.0$    $(1.2)\times 10^{-5} $ and ${\cal{B}}(B^0 \to K^{*0} \tau^- \mu^+)< 8.2$ $(9.8) \times 10^{-6}$  at the 90\% (95\%) confidence level. 
  
\end{abstract}

\vspace*{2.0cm}

\begin{center}
  Submitted to
  JHEP 
 
\end{center}

\vspace{\fill}

{\footnotesize 
% Edit macro in main.tex to keep metadata correct
\centerline{\copyright~\papercopyright. \href{\paperlicenceurl}{\paperlicence}.}}
\vspace*{2mm}

\end{titlepage}

%%%%%%%%%%%%%%%%%%%%%%%%%%%%%%%%
%%%%%  EOD OF TITLE PAGE  %%%%%%
%%%%%%%%%%%%%%%%%%%%%%%%%%%%%%%%

%  empty page follows the title page ----
\newpage
\setcounter{page}{2}
\mbox{~}

%\twocolumn
% %%%%%%%%%%%%% ---------

\renewcommand{\thefootnote}{\arabic{footnote}}
\setcounter{footnote}{0}

%%%%%%%%%%%%%%%%%%%%%%%%%%%%%%%%
%%%%%  Table of Content   %%%%%%
%%%%%%%%%%%%%%%%%%%%%%%%%%%%%%%%
%%%% Uncomment if desired
%\tableofcontents
\cleardoublepage

%%%%%%%%%%%%%%%%%%%%%%%%%
%%%%% Main text %%%%%%%%%
%%%%%%%%%%%%%%%%%%%%%%%%%

\pagestyle{plain} % restore page numbers for the main text
\setcounter{page}{1}
\pagenumbering{arabic}

%\linenumbers

%% This is the main body
\section{Introduction
\label{sec:Introduction}}

Lepton-flavour violation (LFV) has been observed in neutral leptons through the phenomenon  of neutrino oscillations. In  the Standard Model (SM),  LFV is negligible in the charged sector~\cite{Raidal:2008jk}. As a consequence, any observation of a charged LFV decay would provide  clear evidence for physics beyond the SM.   
The recent anomalies observed in some universality tests  of lepton interactions in $b \to s \ell^+ \ell^-$ transitions \cite{LHCb-PAPER-2021-038, LHCb-PAPER-2021-004, LHCb-PAPER-2019-040, LHCb-PAPER-2017-013}   
have led to proposals of many extensions of the SM \cite{Glashow:2014iga, Bhattacharya:2014wla,Becirevic:2016zri, Crivellin:2015era, deMedeirosVarzielas:2015yxm, Becirevic:2016oho, Smirnov:2018ske,  Barbieri:2015yvd, Bordone:2017bld, Bordone:2018nbg, Duraisamy:2016gsd, DiLuzio:2017vat, DiLuzio:2018zxy, Cornella:2019hct}, predicting a significant enhancement of charged LFV decays, including $b \to s \tau \mu$ transitions. 

LHCb has performed searches for lepton-flavour violating $b$-hadron decays into final states with an electron and a muon 
\cite{LHCb-PAPER-2017-031,LHCb-PAPER-2019-022, LHCb-PAPER-2022-008} 
and with a tau and a muon 
\cite{LHCb-PAPER-2019-016, LHCb-PAPER-2019-043}; however, no signal has been observed. The same holds for analogous searches performed by other experiments~\cite{BELLE:2019xld, BaBar:2006tnv, BaBar:2013swg, BaBar:2012azg}.  
The most stringent limit on a $b \to s \tau \mu$ transition is \mbox{$\BR(B_s^0 \to \tau^\pm \mu^\mp) < 3.4 \times 10^{-5}$} at the 90\% confidence level, set by the LHCb experiment \cite{LHCb-PAPER-2019-016}. 

In this article, the first ever search for the charged lepton-flavour violating decay $B^0 \to K^{*0} \tau^\pm \mu^\mp$, not investigated by any prior experiment, 
is presented.   
The $K^{*0}$ meson is reconstructed through its decay into a \Kp and a \pim. 
The final states \OC with the charged kaon and tau having opposite charges and \SC with the charged kaon and tau having the same charge (charged conjugate processes are implied throughout) are treated independently. From a theoretical point of view, these channels could be affected differently by model extensions beyond the SM \cite{Bordone:2019uzc}, and from the experimental point of view they are affected by different background contributions. 
The tau lepton is reconstructed through the decays \tauToPiPiPi or \tauToPiPiPiPiz, representing approximately 14\% of all tau decays.  
 
 The analysis is performed on a data set of proton-proton ($pp$) collisions collected with the LHCb detector at centre-of-mass energies of 7 and 8\tev in 2011-2012 (Run 1) and 13\tev in 2015-2018 (Run 2), corresponding to an  integrated luminosity of 9\invfb.  
 The decay \BdToDDs followed by  \DToKPiPi and \DspToKKPi, with a topology similar to the signal decay and a precisely known branching fraction, is used both as a normalisation channel and a control channel to test the reliability of the simulation and evaluate some systematic uncertainties. 

\section{LHCb detector, trigger and simulation}
\label{sec:detector}

The \lhcb detector~\cite{LHCb-DP-2008-001,LHCb-DP-2014-002} is a single-arm forward
spectrometer covering the \mbox{pseudorapidity} range $2<\eta <5$,
designed for the study of particles containing \bquark or \cquark
quarks. The detector includes a high-precision charged-particle reconstruction (tracking) 
system consisting of a silicon-strip vertex detector surrounding the $pp$
interaction region~\cite{LHCb-DP-2014-001}, a large-area silicon-strip detector located
upstream of a dipole magnet with a bending power of about
$4{\mathrm{\,Tm}}$, and three stations of silicon-strip detectors and straw
drift tubes~\cite{LHCb-DP-2013-003,LHCb-DP-2017-001}
placed downstream of the magnet.
The tracking system provides a measurement of the momentum, \ptot, of charged particles with
a relative uncertainty that varies from 0.5\% at low momentum to 1.0\% at 200\gevc.
The minimum distance of a track to a primary vertex (PV), the impact parameter (IP), 
is measured with a resolution of $(15+29/\pt)\mum$,
where \pt is the component of the momentum transverse to the beam, in\,\gevc.
Different types of charged hadrons are distinguished from one another using information
from two ring-imaging Cherenkov (RICH)  detectors~\cite{LHCb-DP-2012-003}. 
Photons, electrons and hadrons are identified by a calorimeter system consisting of
scintillating-pad and preshower detectors, an electromagnetic and a hadronic calorimeter. 
Muons are identified by a system composed of alternating layers of iron and multiwire
proportional chambers~\cite{LHCb-DP-2012-002}.

The online event selection is performed by a trigger~\cite{LHCb-DP-2012-004}, consisting of a hardware stage,
based on information from the calorimeter and muon systems, followed by a software
stage, which applies a full event reconstruction. In the hardware stage, signal candidates
are required to have at least one high-\pt muon. For the normalisation channel, at least one hadron with high transverse energy is required. 
The software trigger requires a two-, three- or four-track
secondary vertex with a significant displacement from any PV.  At least one charged particle
must have significant transverse momentum and be inconsistent with originating from a PV.
A multivariate algorithm~\cite{BBDT,LHCb-PROC-2015-018} based on kinematic, geometric and lepton identification criteria is used for the identification of secondary vertices consistent with the decay of a \bquark hadron.

Simulation is used to optimise the selection, determine the signal model for the fit
and obtain the selection efficiencies. In the simulation, $pp$ collisions are generated using \pythia 8~\cite{Sjostrand:2007gs} with a specific \lhcb configuration~\cite{LHCb-PROC-2010-056}.  The $B^0$ decay is assumed to proceed according to a uniform  phase-space model. The tau decay is simulated using
the \tauola decay library tuned with BaBar  data \cite{PhysRevD.88.093012}.  The decays of all other
unstable particles are described by \evtgen~\cite{Lange:2001uf}, in which final-state radiation is generated using \photos~\cite{davidson2015photos}.  The interaction of the generated particles with the detector, and its response, are implemented using the \geant toolkit~\cite{Allison:2006ve, *Agostinelli:2002hh} as described in Ref.~\cite{LHCb-PROC-2011-006}.

\section{Event selection}
\label{sec:selection}

The $B^0 \to K^{*0} \tau^\pm \mu^\mp$   candidates are reconstructed by combining six good quality charged tracks, with  $p$ smaller than 110\gevc, $\pt$ larger than 250\mevc, $\eta$  between $2$ and $4.9$. 
Two of the tracks of opposite charge, one compatible with a kaon hypothesis and the other with a pion hypothesis,  are required to form a \Kstarz candidate, with mass within the range  700 to 1100\mevcc and with a good quality vertex. A third track is  identified as a muon. 
The other three remaining tracks, identified as pions, with momentum higher than 2\gevc each, should come from another  vertex.  
These pions form the tau candidate, with charge opposite to that of the muon and with a reconstructed  mass within  0.5 to 2.0\gevcc.    
The tau vertex must have a radial distance between 0.1 and 7\mm and a distance along the $z$-axis larger than 5\mm with respect to the best PV.  

The \Kstarz, the muon and the tau candidates should have a transverse momentum greater than 1\gevc each. They form a  $B^0$ candidate, with a vertex of good quality and sufficiently displaced from any  PV.  
Finally, the $K^{*0}\tau$ mass should be lower than 5\gevcc and the reconstructed $B^0$ candidate mass be within 2 to  10\gevcc. 
A sample of data, with tau and muon of the same charge (called same-sign data in the following), is also selected  as a proxy for backgrounds.

Neutral particles in the tau decays, namely the neutrino and possibly $\pi^0$s, are not explicitly reconstructed. For this reason, the invariant mass of the six tracks, $m_{K^*\tau\mu}$, does not peak at the  $B^0$ meson mass. The  corrected mass $m_{\rm corr} = \sqrt{p_{\perp}^2 + {m^2_{K^*\tau\mu}}} + p_\perp$ is used to recover part of the missing energy, where $p_{\perp}$ is the component of the missing momentum perpendicular to the direction of flight of the $B$ meson~\cite{SLD:2002poq}. 

Backgrounds can be divided into two categories:  combinatorial background, arising from random combinations of tracks; and  physics background due to $b$-hadron decays, which are  partially reconstructed and/or reconstructed with particle mis-identification, called physics background. In particular, a large component of background  involves $D$ mesons with a  decay time compatible with the tau and  decaying  into multiple charged tracks. In order to optimise the rejection of these backgrounds of a different nature,  a multi-stage selection procedure is used. The procedure was developed without looking at candidates in the region where the signal is expected. This region, called the signal region,  is defined  to be $4.6 < m_{\rm corr} < 6.4$\gevcc. Consequently, the region $6.4 < m_{\rm corr} < 18.0$\gevcc is called the upper mass sideband, while the region  $3.0 < m_{\rm corr} < 4.6$\gevcc is called the lower mass sideband. The former is dominated by combinatorial background while the latter contains both combinatorial and physics background.  

The selection  is optimised independently for  Run 1 and Run 2 data-taking periods, to account for possible different background levels due to  different data-taking conditions. The Punzi figure of merit~\cite{Punzi:2003bu},  defined as $\varepsilon/(3/2+{\sqrt{Y_b}})$, is used for identifying the optimal selection requirement, apart from the vetoes. Here $\varepsilon$ is the efficiency of the signal selection, measured with simulation, and  $Y_b$ is the background yield in the signal region, estimated differently depending on the selection step. 

The first stage of the selection is based on a multivariate discriminant  exploiting the differences between the topologies of the signal decays and the combinatorial background. A boosted decision tree (BDT)~\cite{Breiman}, using the AdaBoost algorithm from the TMVA package~\cite{Speckmayer:2010zz},  combines information from: the $\chi^2$ of the vertices of the \Bd, \Kstarz and tau candidates; the $\chi^2$ of the flight-distances of the \Bd and the tau candidates. 
The BDT  is trained using  simulated signal samples  and the upper sideband of the same-sign data sample as a background proxy.  
A \kfolding\ approach \cite{Bevan:2017stx} is used  to exploit the training samples without biasing the output of the classifiers.  For the threshold optimisation, the background yield in the signal region is extrapolated from a fit with a decreasing exponential function to the candidate  $m_{\rm corr}$  distribution in the region 11 to  18\gevcc, dominated by combinatorial background.  

The second stage is a multivariate selection dedicated  to the rejection of charmed mesons mis-identified as  tau leptons.  The decay of a tau into three charged pions and a neutrino occurs mostly through the $a_1^\pm$ resonance, which in turn decays into a $\rho^0$ and a $\pi^\pm$ meson. In order to exploit the   kinematic properties of this decay chain,   the minimum and maximum  of the momenta of the pions  from the tau candidate and the  masses of pion pairs with zero combined charge are combined into a BDT. The BDT is trained using a \kfolding\ approach, with  \osBToKstTauMu  simulation  and  inclusive  \bbbar simulated events reconstructed as \osBToKstTauMu as a proxy for the background. Particle identification requirements and \Kstarz kinematic constraints are removed, so that particles  from charmed mesons in the \bbbar sample can form tau and \Kstarz candidates. 
For the threshold optimisation, 
the background yield is extrapolated from a fit to the \Bd candidate mass distribution in the 3 to 18\gevcc mass range, excluding the signal region. The fit model used is the sum of a Crystal-Ball function~\cite{Skwarnicki:1986xj}, modelling the lower mass sideband, dominated by partially reconstructed events, and an exponential function, with a  slope  primarily determined by events in the upper mass sideband.  
The reliability of this background yield estimation  is verified on the same-sign data, where the prediction can be compared to the observed number of candidates in the signal  region. 

The third selection step consists of  requirements on the muon, kaon and pion particle-identification variables.  A three-dimensional scan over the possible thresholds of the three selection criteria is performed, and a configuration  maximising  the Punzi figure of merit is chosen. 
For this optimisation, the background yields evaluated in the previous step are scaled  by the rejection of the particle-identification requirements, as estimated from the same-sign sample within the signal region. 

  The search sensitivity is further  increased by  exploiting the discriminating power of the mass of \Kstarz candidates, required to be within 856 to 946\mevcc.  
In addition, the mass of the tau candidate is calculated using the known mass of the pion~\cite{PDG2020} and  considering only the  pions and neutrino momentum components orthogonal to the \Bd direction of flight.    The neutrino component is measured as the difference between the momentum orthogonal components of the  $K^{*0}\mu$ system and the three pions system. This mass is required to be within 789 to 1900\mevcc.  
Both mass regions have been optimised using the Punzi figure of merit,   as in the previous step. 
A particle selected from a partially reconstructed decay is oftentimes surrounded by other particles which are not used to reconstruct the candidate. 
To exploit this feature, a Fisher discriminant~\cite{fisher36lda} is built, using the \kfolding\ procedure and based on the following variables evaluating the particle isolation: 
the logarithm of the smallest variation of the \Bd, the \Kstarz or the tau  vertex $\chi^2$ when adding  one or two tracks to them, independently; 
and the invariant mass of all the particles forming the vertex under the hypotheses of such additions; the number of tracks compatible with the aforementioned vertices (\ie whose addition keeps the overall $\chi^2$ below 9). The  Fisher discriminant is chosen for  its stability in the training procedure, despite the small size of the training samples available at this point of the selection. Signal simulation and, for the background, same-sign data  that  fall into the signal mass region, are used for the training and  for the optimisation of the threshold. 

A powerful variable to reject the physics backgrounds is the tau flight-distance $\chi^2$; it is already used to reject the combinatorial background, which tends to have larger values  than the signal. It also offers  residual discriminating power against physics backgrounds from charmed mesons, for example, $B^0 \to D^{*-} \mu^+ \nu$ decays for the \OC decay channel. Given the different nature of the physics backgrounds, the thresholds applied to the \OC and \SC channels are optimised separately. 

The remaining physics backgrounds are studied with events rejected by the first BDT and within the region $4.7<m_{\rm corr}<5.7$\gevcc.  
Mass distributions from possible final state particle combinations are used to define vetoes that remove the observed resonant structures. 

For the \OC case, an important  physics background  comes from the decay  $B^0 \to D^{*-} \mu^+ \nu$, with $D^{*-}\to \Dzb \pi^-$ and $ \Dzb  \to K^+ \pi^- \pi^+ \pi^-$. Three of the four pions  can be mis-reconstructed as a tau candidate, while the remaining pion and  kaon are used to mis-reconstruct  a  $K^{*0}$ candidate.   
To reduce the backgrounds from \Dz mesons, events with $m(K^+ \tau^-)$ or $m(K^{*0}\pi^-\pi^+)$ within 60\mevcc from the known \Dz mass  are removed. This requirement additionally rejects most of the \Dstarz contributions. However,  considering the minimal  impact  on the signal efficiency, events with $m(K^{*0}\tau^-)-m(K^{*0}\pi^-\pi^+)$ or  $m(K^{*0}\tau^-)-m(K^+\tau^-)$  between 135.5 and 155.5\mevcc are also rejected. 
Masses accounting for a possible mis-identification of the muon as a pion have been checked, but no significant contributions from charmed mesons have been observed.    

For \SC decays, a significant source of physics background at this stage of the selection comes from the decay $B^0 \to D^{*-} \tau^+ \nu_{\tau}$, with $D^{*-}\to \bar{D^0} \pim$,  $ \bar{D^0} \to K^+ \mu^- \bar{\nu_{\mu}}$ and $\tau^+ \to \pi^+ \pi^- \pi^+ \bar{\nu_{\tau}}$.   This background is rejected by selecting only events with  $m(K^+\mu^-)>1885$\mevcc.  

After the selection procedure described above, there is never more than one $B^0$ candidate selected per event.

\section{Normalisation channel}
\label{sec:normalisation}
The normalisation channel  \BdToDDs, with \DToKPiPi and \DspToKKPi, is reconstructed using six good-quality tracks, with $2< p<$ 110\gevc, $\pt>250$\mevc, $2<\eta<4.9$ and having associated deposits in the RICH detectors.  Three of these tracks, one  identified as kaon and the other two as pions, are used to form a $D^-$ candidate with  a  displaced vertex of good quality and an invariant mass lying within the range 1750 to 2080\mevcc. Analogously,  one track identified as pion and  other two tracks  identified as kaons form the $\Ds$ candidate, with  a displaced vertex of good quality and an invariant mass within the range 1938 to 1998\mevcc. Both the $D^-$ and $D_s^+$ candidates are required to have  $\pt> 1$\gevc. 

Combinatorial background is rejected by  a  BDT  exploiting the same topological variables used in the selection of the signal channel, but with  two flight-distance significances, one for each $D$ meson. The BDT is trained using  simulated \BdToDDs decays  and \BdToDDs candidates with a mass  exceeding 5400\mevcc as a background proxy. A \kfolding\ procedure is applied. 
Considering that a significant signal is expected for the normalisation mode, the selection optimisation is based on the figure of merit  $Y_s/\sqrt{Y_s+Y_b}$,  where $Y_s$ and $Y_b$ are the yields of the signal and background, respectively. These yields are  determined from fits in the \Bd mass range 5150 to  5400\mevcc, using a Gaussian model to parameterise the  signal component and a decreasing exponential function to describe  the background.  

The same particle identification requirements used for selecting the signal are applied on  kaons and pions. For each charmed meson, the difference between its reconstructed mass and known mass~\cite{PDG2018} must be less than 20\mevcc.  

\begin{figure}[htbp]
	\centering
		\includegraphics[width= .495\linewidth]{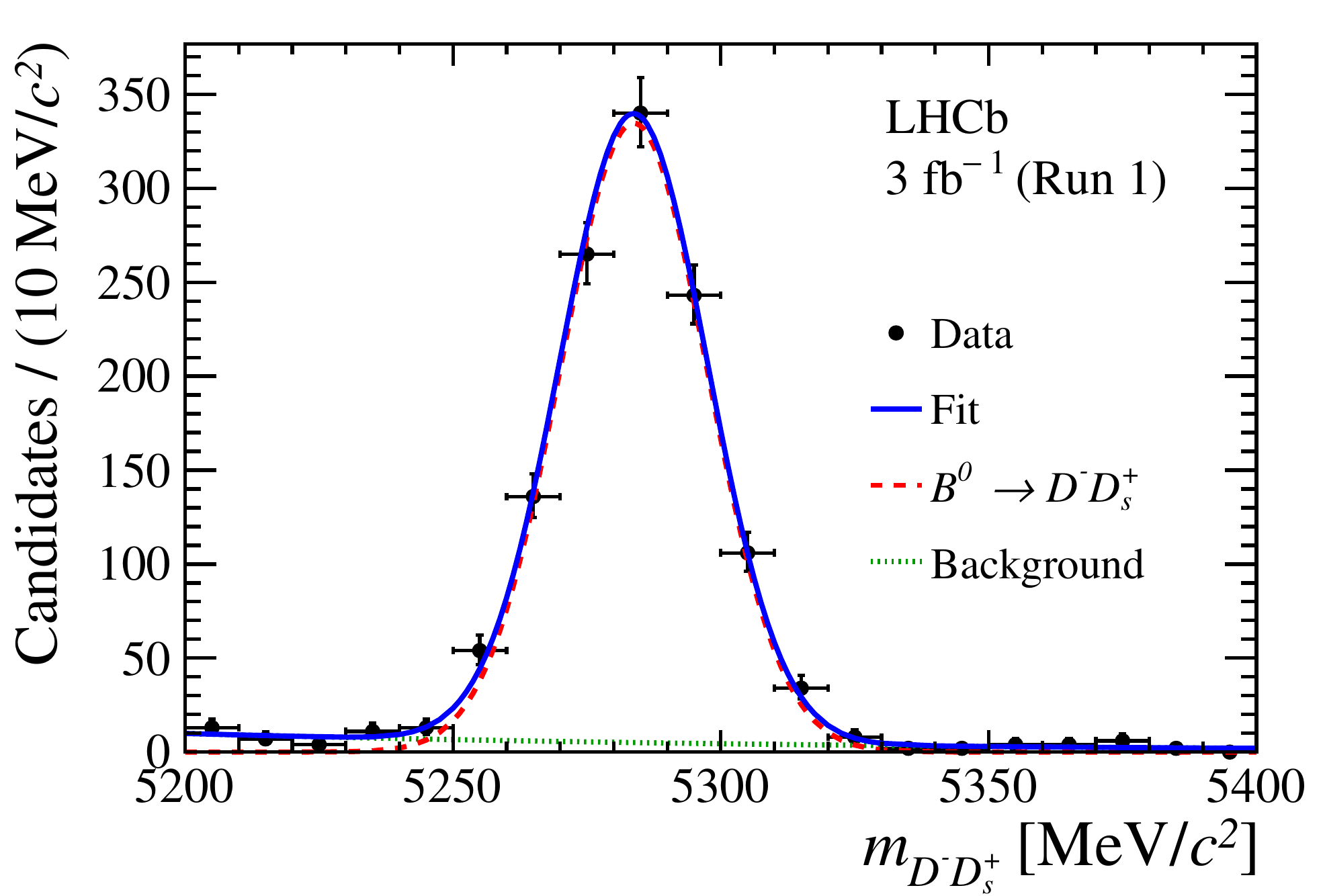}
		\includegraphics[width= .495\linewidth]{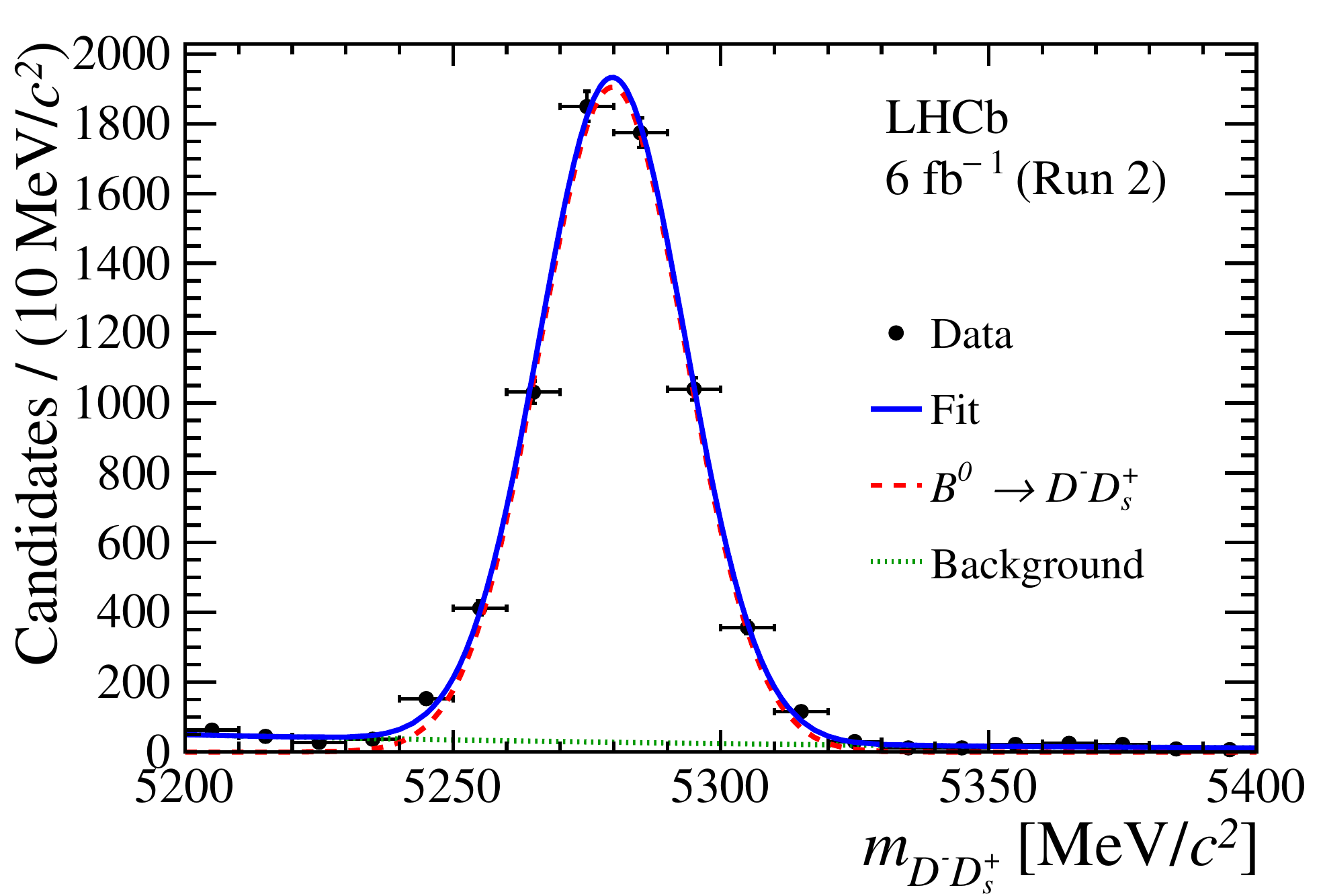}
	    \caption[Normalisation channel fit]{Mass distributions of selected $D^- D_s^+$  
	    %\BdToDDs 
	    candidates in (left) Run 1 and (right) Run 2 data, with the fit overlaid (blue solid line).  The red dashed line is signal, while the green dotted line is background.
              }
	    \label{fig:dds_fit}
\end{figure}

Finally, the normalisation channel yields are measured from fits of the \Bd mass distribution   
for each year of data taking.  
The global fits  for Run 1 and Run 2  data, leading to yields of $1155\pm 35$  and $6516\pm 84$, respectively, are shown in Fig.~\ref{fig:dds_fit}.

\FloatBarrier

\section{Determination of efficiencies}
\label{sec:effs}

The efficiencies for the different steps of the selection chain are evaluated separately for each year of data taking. 
With the exception of the particle identification efficiency,  they are estimated with simulation. 
The reliability of the simulation 
is checked using \Bd, \Dm and \Ds mesons from the normalisation channel \BdToDDs (with \DToKPiPi and \DspToKKPi)  as proxies for two- and three-prong particle vertices, selected with the additional  requirement that the \Bd candidate mass lies within  $\pm 50$\mevcc from its nominal value.
All the variables used in the multivariate selections are found to be well described by simulation. Residual discrepancies are assessed as a systematic uncertainty.

The particle identification variables are not perfectly described in simulation. Therefore, the particle identification efficiency is evaluated using data calibration samples.  
A pure sample of pions and kaons is obtained from the decay $\decay{\Dstar^+}{(\decay{\Dz}{\kaon\pion})\pion^+}$ while muons are selected from the decay $\decay{\jpsi}{\mumu}$, without relying  on PID selection criteria~\cite{LHCb-DP-2018-001}.  The efficiencies of the PID requirements can then be computed in intervals of the kinematic variables of the final-state-particles momentum and pseudorapidity~\cite{LHCb-PUB-2016-021}. Intervals are chosen making a compromise between the stability of the efficiency within the interval and statistical uncertainty.  The event occupancy, parameterised by the number of tracks per event, is also taken into account,  relying on  same-sign events in the case of signal and on the observed distribution for the normalisation channel. 

\begin{figure}[htbp]
\centering
\includegraphics[width=.48\linewidth]{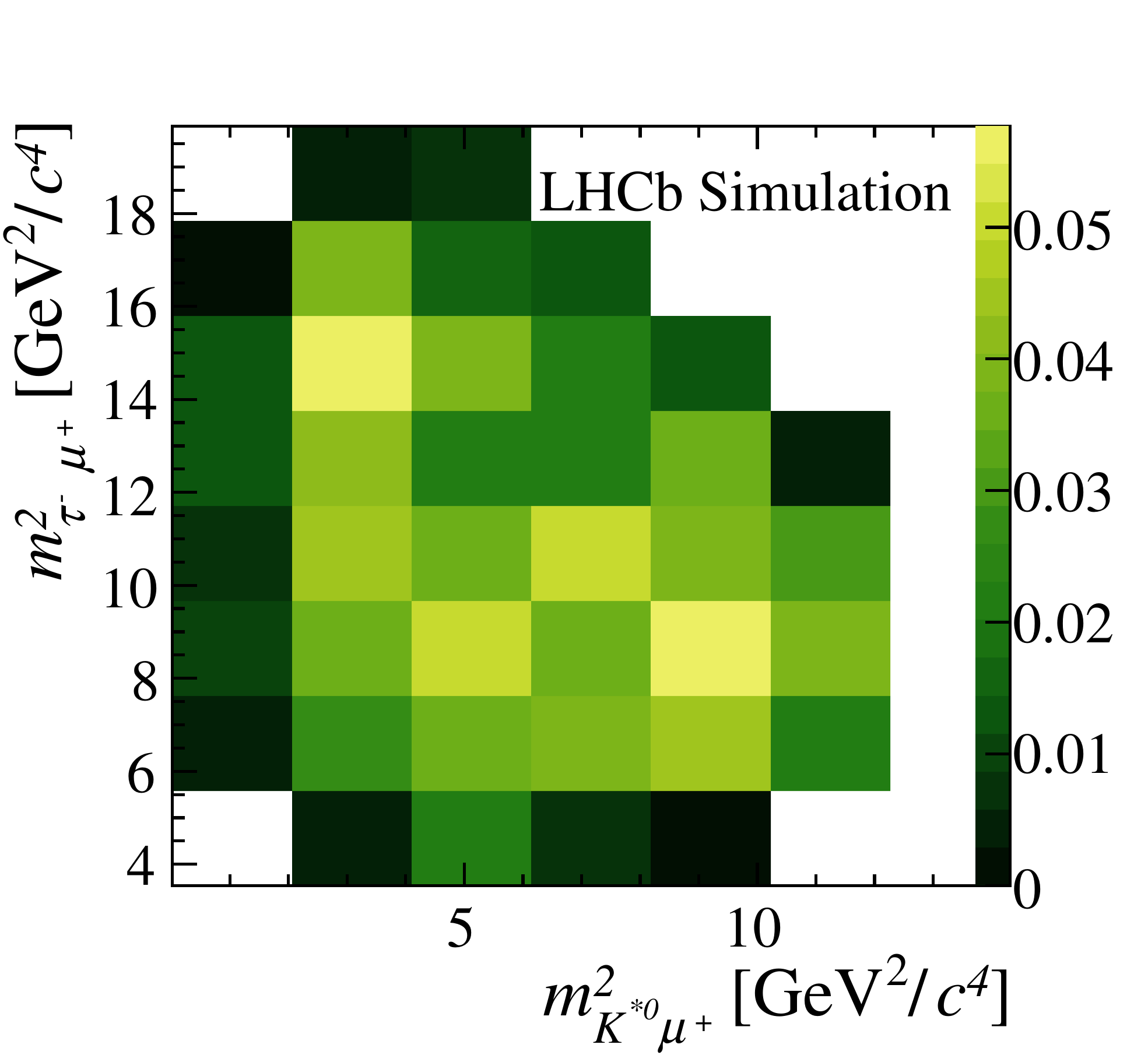}
\includegraphics[width=.48\linewidth]{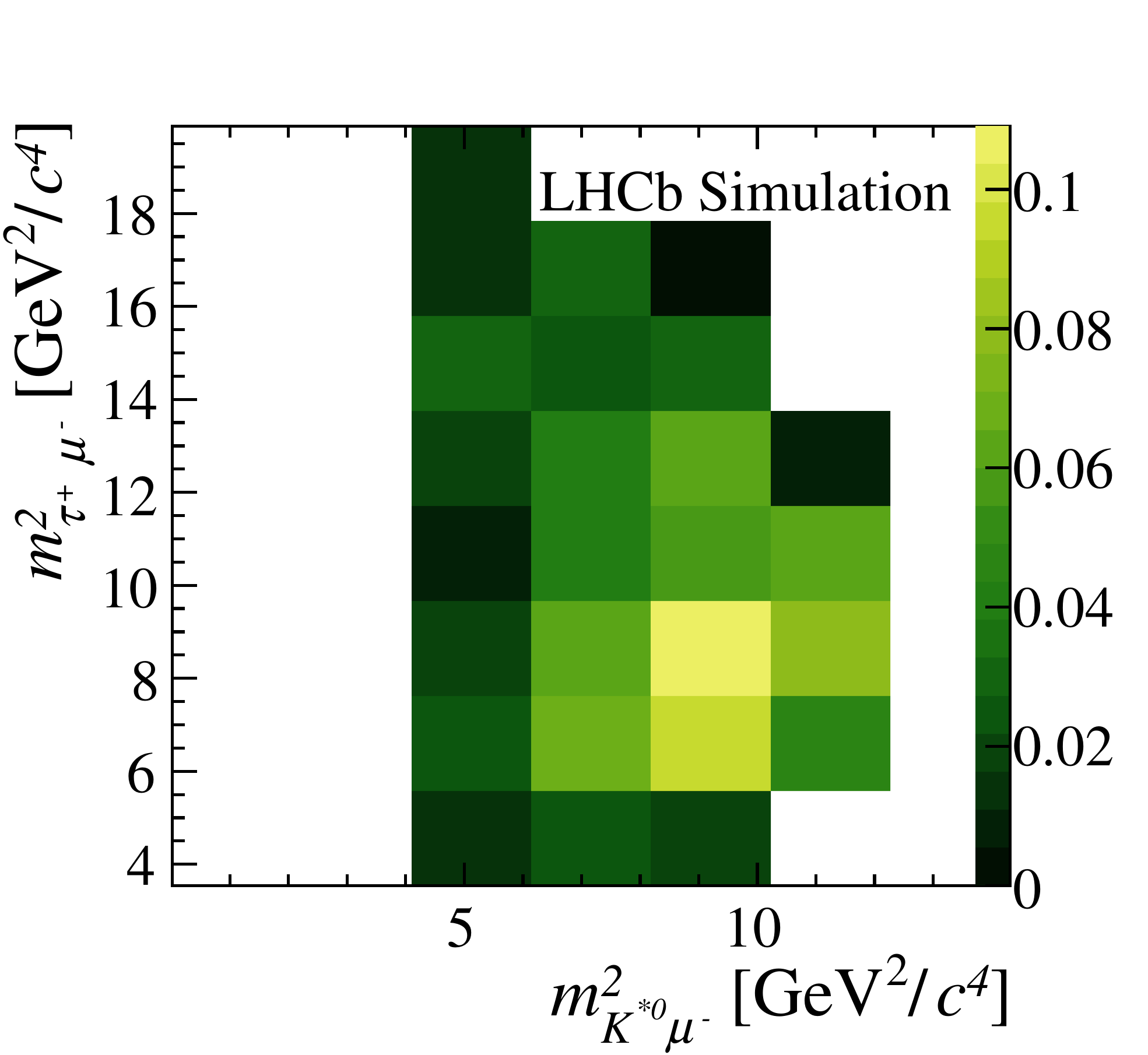}
	    \caption[Distribution of the efficiencies]{ Efficiency as a function of $m^2_{K^{*0}\mu^{\pm}}$ and $m^2_{\tau^{\mp}\mu^{\pm}}$  for the (left) \mbox{\OC}  and (right) \mbox{\SC}  cases, normalised to unity. The selection requirement $m(K^+\mu^-)>1885$\mevcc for the \SC decay implies zero efficiency  for \mbox{$m^2_{K^{*0}\mu^-}<4.1$ \gevgevcccc} on the right plot.}
	    \label{fig:eff}
\end{figure}

The distribution of the  selection efficiency as a function  of the kinematic variables $m^2_{K^{*0}\mu^{\pm}}$ and $m^2_{\tau^{\mp}\mu^{\pm}}$, normalised to unity,  is shown in Fig.~\ref{fig:eff}. These distributions can be used to re-cast the results for models having \OC and \SC decay kinematics different from the uniform distribution in the phase space  used for the signal simulation in this analysis.

\section{Strategy to fit the corrected mass distributions}
\label{sec:fitMass}
An extended maximum likelihood fit to the $m_{\rm corr}$ distributions of the selected events is performed independently for the \OC and \SC decay channels.  For each channel the distribution is described by a 
function $P_{\rm{tot}}$ which is the sum  of three components,   $P_{\rm{tot}} =  Y_{\tau_{3\pi}}  P_{\tau_{3\pi}} +  Y_{\tau_{3\pi\pi^0}}  P_{\tau_{3\pi\pi^0}} + Y_{\rm{bkg}}  P_{\rm{bkg}}$.  The function $P_{\tau_{3\pi}}$ models the dominant signal contribution with \tauToPiPiPi 
and $Y_{\tau_{3\pi}}$ indicates the corresponding yields. Analogously,  $P_{\tau_{3\pi\pi^0}}$  models the subdominant  signal contribution with  \tauToPiPiPiPiz 
and  $Y_{\tau_{3\pi\pi^0}}$ represents the corresponding yields.  Finally, $P_{\rm{bkg}}$ models the background, with yields $Y_{\rm{bkg}}$.

The signal components $P_{\tau_{3\pi}}$ and   $P_{\tau_{3\pi\pi^0}}$ are parameterised independently, using  double-sided Crystal-Ball 
functions (DSCB),  
\ie a crystal ball with an additional tail  on the right side. The parameters are determined and fixed from fits to the simulated signal samples.   
The  yields for the dominant signal contribution are expressed, for the \OC channel and separately for Run 1 and Run 2, as 
\begin{equation}
Y_{\tau_{3\pi}}^{\rm{run}} =   
\frac{ \BR(\OC) \cdot  \BR(\KstToKPi)  \cdot \BR (\tauToPiPiPi)}{  \BR(\BdToDDs) \cdot \BR(\DToKPiPi)  \cdot \BR(\DspToKKPi) } \cdot 
\sum_{y \in {\rm{run}}}
{\left(\frac{\varepsilon Y_{\rm{N}}  }{\varepsilon_{\rm{N}}}\right)_{y}}.
\end{equation}
 Analogous expressions hold for the \SC channel, and for the subdominant signal components. Here, $\epsilon$ represents the signal efficiency, while $Y_{\rm{N}}$ and $\epsilon_{\rm{N}}$ indicate the yields and the efficiency for the normalisation channel. The  ratio $\varepsilon Y_{\rm{N}}  /\varepsilon_{\rm{N}}$ is calculated for each year in the run, as indicated by the $y$ index.  
 The parameter of interest $\BR(\OC)$ is in common among the  yields of the dominant and subdominant signal contributions, as well as among runs of data taking, and is therefore fitted simultaneously. A fit bias on the measured signal branching ratio, evaluated on background-only pseudoexperiments, is at the level of $~10^{-7}$ and is subtracted. The other branching fractions are taken from Ref.~\cite{PDG2020}. Apart from the signal branching fraction, all the  parameters are constrained with Gaussian functions to account for the systematic uncertainties described in the next section. 

The functional form used to parameterise the background shape is also a DSCB function. The background yields are free to vary in the fit, while the shape parameters are constrained to the value determined on a control sample obtained by loosening the combinatorial multivariate selection. Signal contamination in the control regions 
is estimated to be less than 5\% of the total signal, assuming a branching fraction of $10^{-5}$.

\section{Systematic uncertainties}
\label{sec:systematics}
For the determination of the limit, some branching fractions are used as external inputs. A systematic uncertainty is assigned imposing Gaussian constraints on their values, taken from the PDG\cite{PDG2020}.

The normalisation procedure involves ratios of signal and normalisation efficiencies, for which  systematic uncertainties cancel.  However, some efficiencies do not cancel in the ratio and are assessed as follows. 

The uncertainty related to the limited size of the simulation samples used to determine the efficiencies is included as part of the statistical  uncertainty.  

The ratio between the signal and the normalisation tracking efficiencies is determined from simulation. Possible differences in data with respect to simulation will be of the same order in the numerator and denominator for five out of the six particles in the final state, and will cancel. 
Because the sixth hadron in the normalisation channel interacts differently with the material with respect to the muon in the signal channel, a 1.4\% uncertainty   
is assigned to the tracking efficiency ratios~\cite{LHCb-DP-2013-002}.  

The systematic uncertainty on the determination of the PID selection efficiency accounts for: the limited size of simulation and calibration samples; the choice of the $p$, $\eta$ and number of tracks interval sizes; and the use of the $sPlot$ technique~\cite{Pivk:2004ty} to extract the signal yield in the control samples.

Although the simulation of the variables used in the multivariate classifiers is validated with the normalisation channel, small residual discrepancies  could  affect the evaluation of the classifier efficiencies. To account for this, the \BdToDDs candidates with a mass within $50$\mevcc of the known mass~\cite{PDG2020} of the \Bd meson are selected from both data and simulation, providing a high purity sample on which the classifiers used to select the \BToKstTauMu signal are applied. The variables of the classifiers relative to the  tau  are applied to the \Dm meson  and analogously those of the \Kstarz are applied to the \Ds meson. For each classifier,  the requirement giving  an efficiency on the \BdToDDs simulation  sample equivalent to that obtained by the  requirement on the simulated \BToKstTauMu sample is determined. The relative systematic uncertainty is  the absolute value of the difference in efficiencies of this requirement between \BdToDDs simulation and data, divided by the \BdToDDs simulation efficiency.

Over the entire data taking period,  changes were applied to the hardware muon trigger, in particular regarding the requirement on the transverse momentum of the muon. To account for these changes,  
 the hardware muon trigger efficiency evaluated from simulation is multiplied by a correction factor derived as the ratio between the efficiency from data and simulation for the $B^\pm  \to \jpsi K^\pm$ control sample. These corrections are functions of the muon transverse momentum. The difference between the efficiency resulting from this weighting procedure and the efficiency from simulation is taken as a systematic uncertainty. The overall effect is less than 1\%. 

The hardware hadron trigger, used in the selection of the normalisation channel,   is less well reproduced in simulation. 
A data-driven efficiency estimation is provided by  the number of events triggered simultaneously by any  muon in the  event and  one of the hadrons in the \BdToDDs decay, divided by the number of events triggered by any muon in the event. The absolute difference with the efficiency  estimated on the \BdToDDs simulation is assumed as a  systematic uncertainty.

The systematic uncertainty associated to the fit model used for measuring the normalisation channel yields ($Y_n$) is assessed via pseudoexperiments, generated using  a kernel estimation of the  \BdToDDs mass distribution and subsequently fit using the nominal normalisation channel fit model.  The relative systematic uncertainty is the ratio between the fitted mean of the yields with respect to the mean of the generated values, and is 1.8\% for Run~1 and 1.7\% for Run~2.

The parameters of the DSCB function describing the background are Gaussian constrained to the values  determined by the fit of the background control region, obtained by loosening the combinatorial multivariate selection. Alternative control regions are defined by choosing different requirements on the combinatorial multivariate discriminant. The mean value and width of the DSCB are Gaussian constrained to half  the maximum spread of the results obtained  from fits on the alternative control regions, if this is larger than the parameter uncertainty on the default control region.    
The  systematic uncertainty for the choice of the DSCB functional form  is assessed  by generating samples using  a kernel estimation of background control samples,  and fitting with the default DSCB function. 
A bias term is added to each of the signal branching ratios, and its value is constrained to a Gaussian function with a mean of zero and a  width set to the average value of the bias measured on pseudoexperiments, which is at the level of $6\times 10^{-7}$.  

The effect of the systematic uncertainties is summarised in Tab.~\ref{tab:syst}, where the increase of the observed upper limit when applying each of the systematic uncertainties in addition to the others is shown.   
The dominant systematic effect comes from the uncertainty on the arbitrary choice of the control region.

\begin{table}[t]
	\centering
	\caption{Relative increase (in \%) of the observed upper limit when applying each of the systematic uncertainties.}
	\label{tab:syst}
	\renewcommand\arraystretch{1.3}
	%\footnotesize
	\begin{tabular}{l|cc}
		\textbf{Systematic effect} & \multicolumn{2}{c}{\textbf{Limit increase [\%]}} \\
		 &  \textbf{\OC} & \textbf{\SC} \\
 
		\hline
 
          Input branching fractions & 4 & 3 \\
         Efficiencies & 2 & 1 \\
         Normalisation yields & 1 & 1 \\
         Background control region choice & 18 & 26 \\
                  Background analytical shape & 1 & 1 \\
	\end{tabular}
\end{table}
\section{Results}
\label{sec:results}

\begin{figure}[htbp]
	\centering
	\includegraphics[width= .48\linewidth]{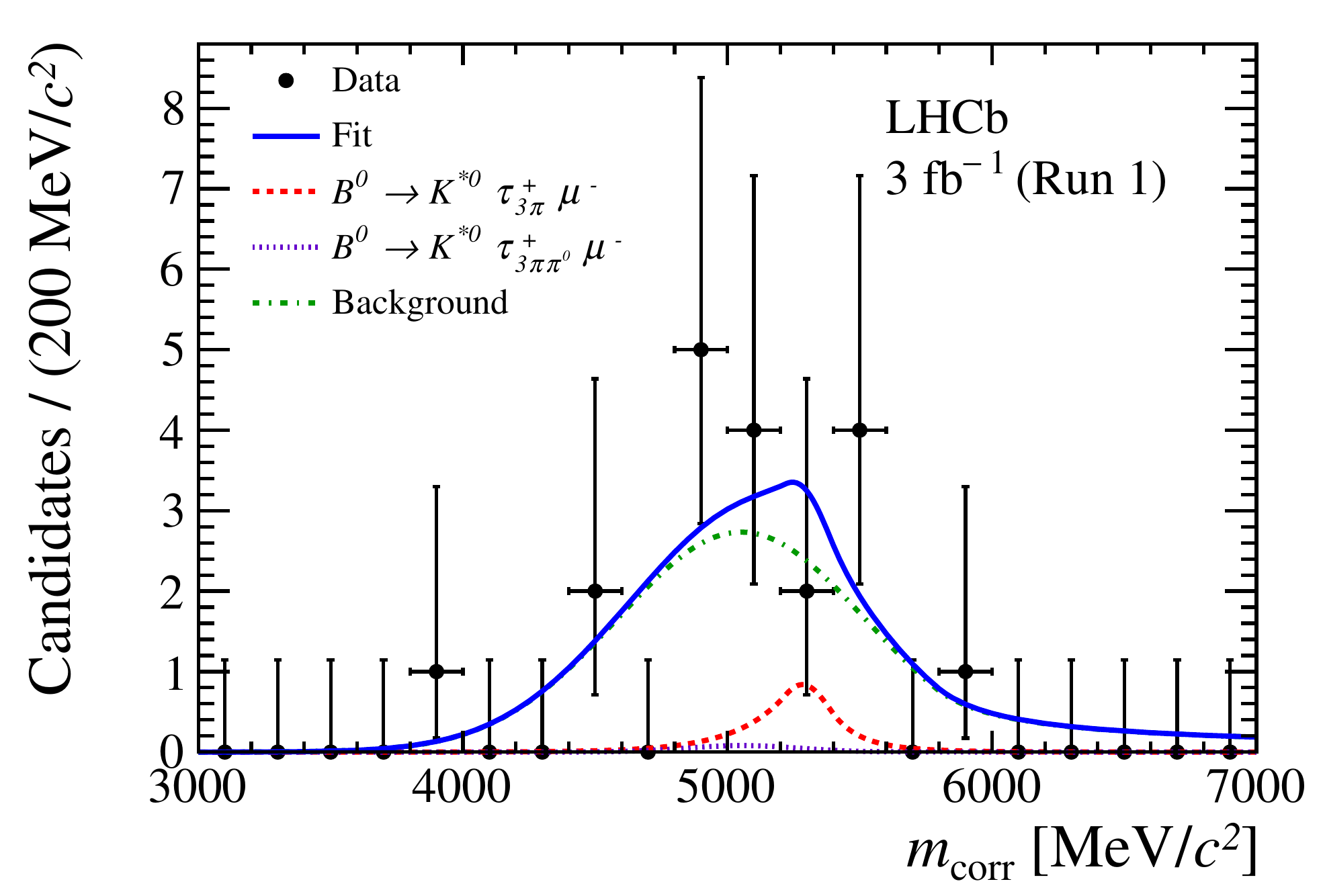}
	\includegraphics[width= .48\linewidth]{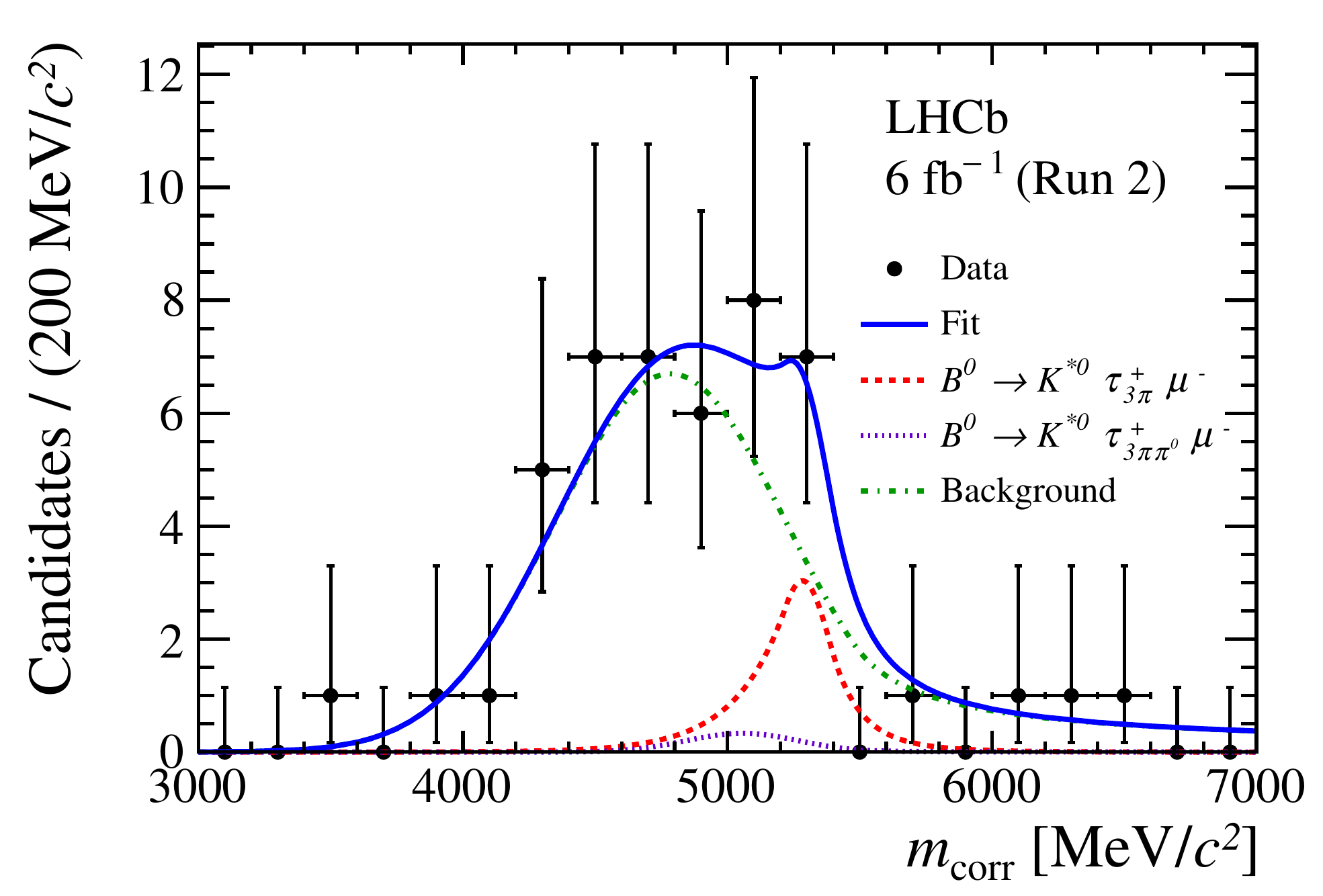}
	    \caption[Fit oc]{
	    Distribution of the corrected mass $m_{\rm corr}$  of selected \SC  candidates in (left) Run 1 and (right) Run 2 data, with the simultaneous fit overlaid (blue solid line).  The red dashed line is the dominant signal component with $\tau^+ \to \pi^+ \pi^- \pi^+ \bar{\nu_{\tau}}$;  the violet dotted line is the subdominant signal component with $\tau^+ \to \pi^+ \pi^- \pi^+ \pi^0 \bar{\nu_{\tau}}$, extremely small and consequently barely visible; and the green dash-dotted line is the  background.
	    }
	    \label{fig:sc_fit}
\end{figure}

\begin{figure}[htbp]
	\centering
	\includegraphics[width= .48\linewidth]{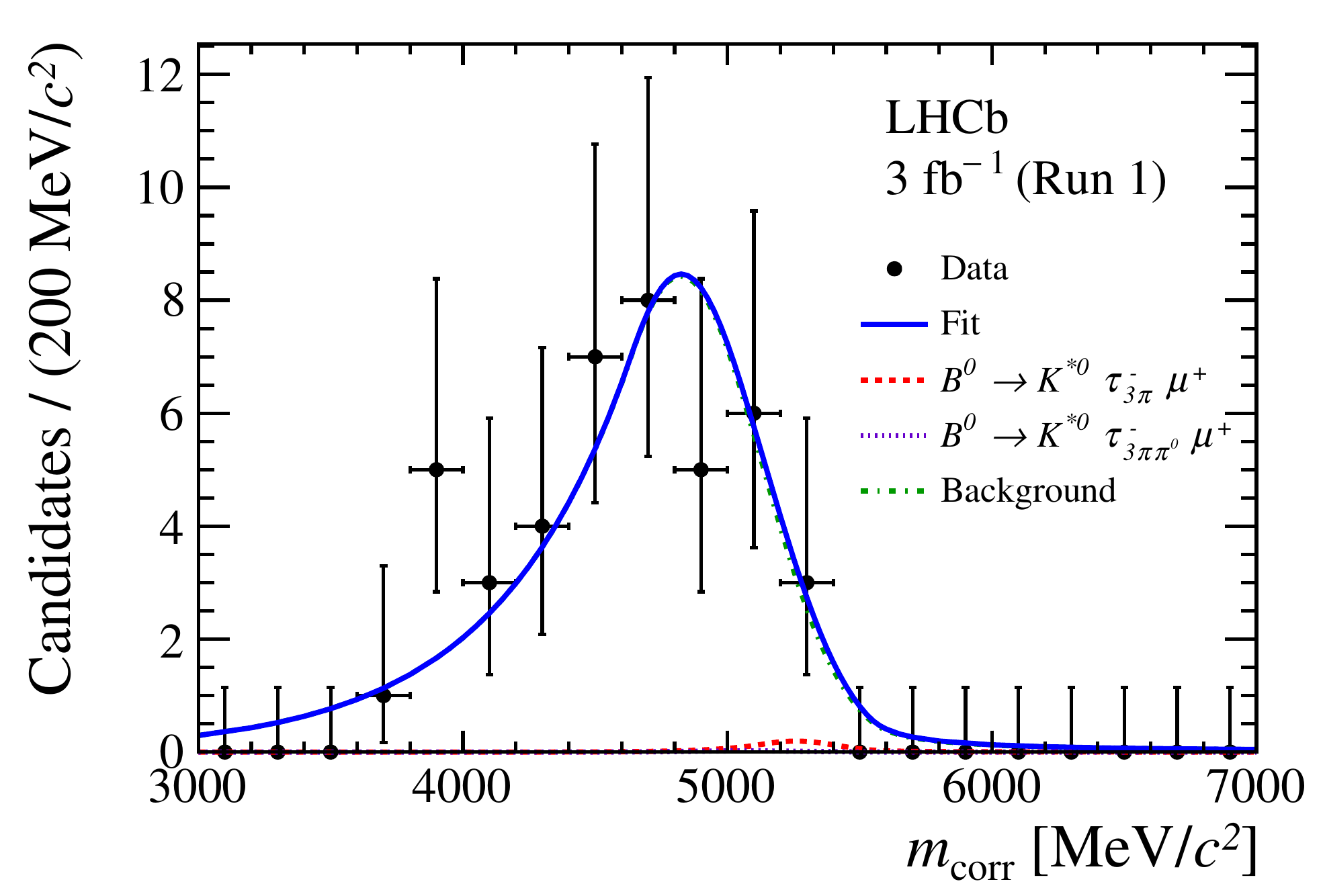}
	\includegraphics[width= .48\linewidth]{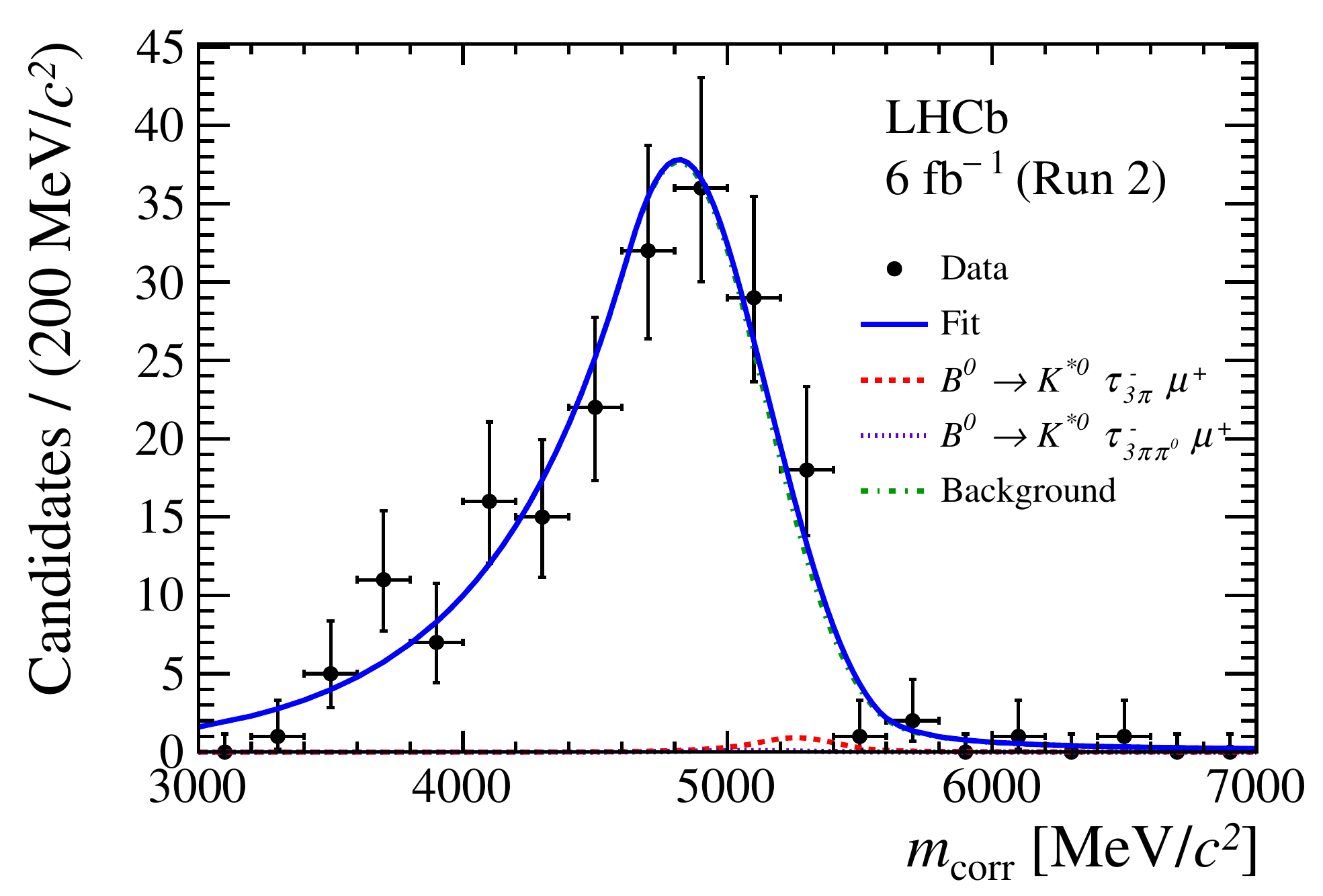}
	    \caption[Fit oc]{
	    Distribution of the corrected mass $m_{\rm corr}$  of selected \OC  candidates in (left) Run 1 and (right) Run 2 data, with the simultaneous fit overlaid (blue solid line). 
	    The red dashed line is the dominant signal component with $\tau^- \to \pi^- \pi^+ \pi^- {\nu_{\tau}}$;  the violet dotted line is the subdominant signal component with $\tau^- \to \pi^- \pi^+ \pi^- \pi^0 {\nu_{\tau}}$, extremely small and consequently barely visible; and the green dash-dotted line is the  background.
	    }
	    \label{fig:oc_fit}
\end{figure}

\begin{figure}[htbp]
	\centering
	\includegraphics[width= .48\linewidth]{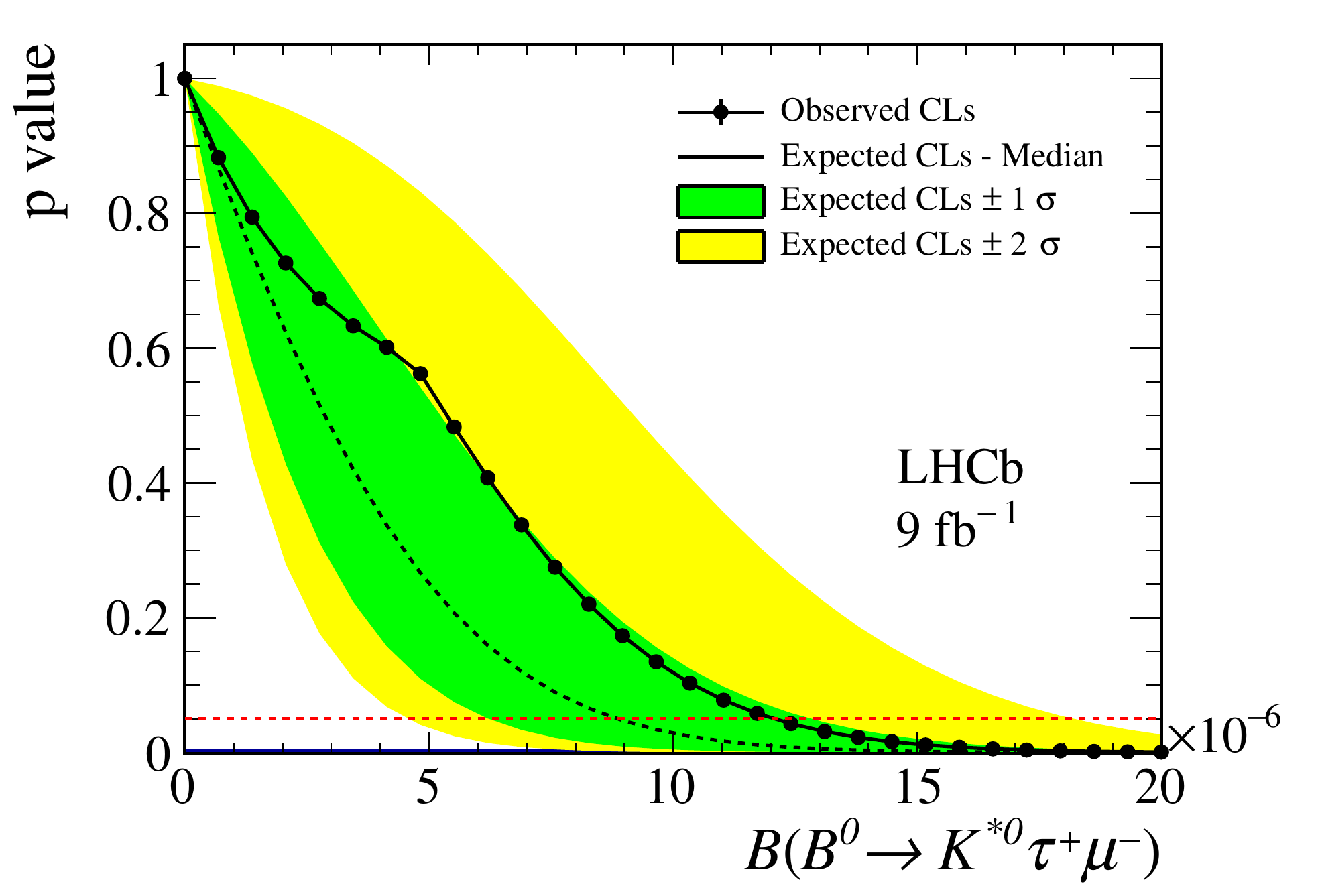}
	\includegraphics[width= .48\linewidth]{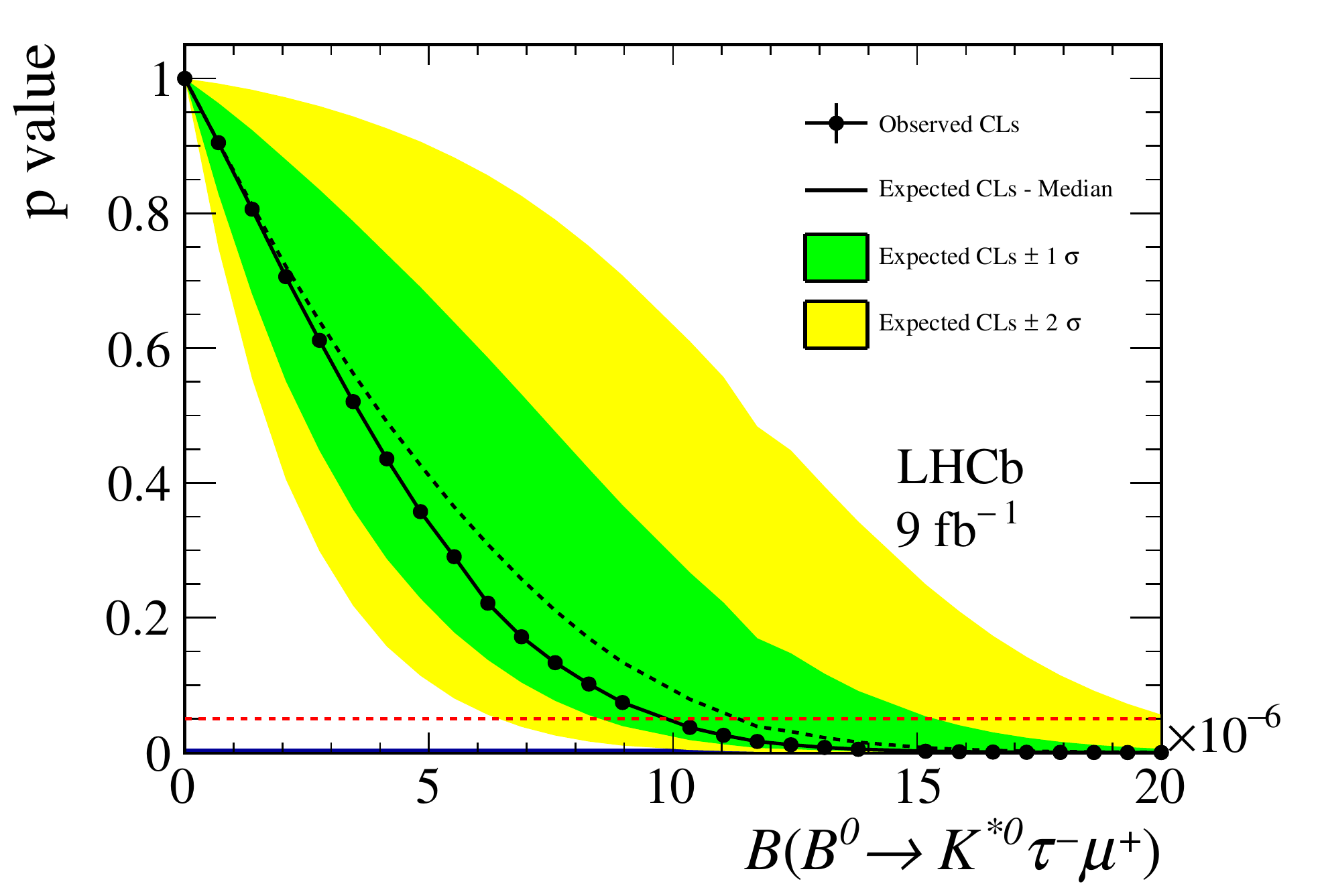}
	    \caption[Fit oc]{The expected and observed p-values derived with the $CL_s$ method as a function of  the  (left) \SC  and (right) \OC  branching fraction. The red line corresponds to the 95\% CL.  }
	    \label{fig:limits}
\end{figure}

The results of  the extended maximum likelihood fit to the $m_{\rm corr}$ distributions are shown in Figs.~\ref{fig:sc_fit} and~\ref{fig:oc_fit} for  \SC and \OC, respectively. No significant signal contribution is observed.  Therefore, upper limits on the branching fractions are set  via the $\mathrm{CL_s}$ method \cite{CLs, Cowan:2010js}, using the asymptotic approximation: \mbox{$\BR(\SC)< 1.0$    $(1.2)\times 10^{-5} $} and \mbox{$\BR(\OC)< 8.2$ $(9.8) \times 10^{-6}$}  at the 90\% (95)\% confidence level,  as shown in Fig.~\ref{fig:limits}.


\section{Conclusion}
\label{sec:conclusion}

The first search for the lepton-flavour violating decays \osBToKstTauMu 
has been performed on a sample of proton-proton collision data, collected with the LHCb detector at centre-of-mass energies of 7, 8 and 13\tev  between 2011 and 2018, corresponding to an integrated luminosity of 9\invfb. No significant signal is observed, and upper limits on the branching fractions are set: \mbox{$\BR(\SC)< 1.0$    $(1.2)\times 10^{-5} $} and \mbox{$\BR(\OC)< 8.2$ $(9.8) \times 10^{-6}$}  at the 90\% (95\%) confidence level. These results assume a uniform distribution of the signal events within the phase space accessible to the $K^{*0}$, tau and muon. They are currently the most stringent upper limits on  $b \to s \tau \mu$ transitions.

\addcontentsline{toc}{section}{References}
%\setboolean{inbibliography}{true}
\bibliographystyle{LHCb}
\bibliography{main,standard,LHCb-PAPER,LHCb-CONF,LHCb-DP,LHCb-TDR}

\newpage

\end{document}